\documentclass[letter]{aa}

\usepackage{graphicx}
\usepackage{epsfig}
\usepackage{natbib}
\usepackage{txfonts}
\usepackage{balance}
\usepackage{hyperref}

\newcommand{\COBOLD}{{\tt CO$^5$BOLD}}

\newcommand{\LHD}{{\tt LHD}}

\newcommand{\Linfor}{{\tt Linfor3D}}

\newcommand{\ATLAS}{{\tt ATLAS9}}
\newcommand{\MULTI}{{\tt MULTI}}

\newcommand{\naofe}{\ensuremath{\left[\mathrm{Na}/\mathrm{Fe}\right]}}
\newcommand{\oofe}{\ensuremath{\left[\mathrm{O}/\mathrm{Fe}\right]}}
\newcommand{\liona}{\ensuremath{\left[\mathrm{Li}/\mathrm{Na}\right]}}
\newcommand{\lioo}{\ensuremath{\left[\mathrm{Li}/\mathrm{O}\right]}}
\newcommand{\naoo}{\ensuremath{\left[\mathrm{Na}/\mathrm{O}\right]}}
\newcommand{\ALi}{\ensuremath{A({\rm Li})}}
\newcommand{\AO}{\ensuremath{A({\rm O})}}
\newcommand{\ANa}{\ensuremath{A({\rm Na})}}
\newcommand{\Teff}{\ensuremath{T_{\mathrm{eff}}}}

\begin{document}

\title{Galactic globular cluster 47 Tucanae: new ties between the chemical and dynamical evolution of globular clusters?}

   \subtitle{}

\author{
        A.~Ku\v{c}inskas\inst{1}
        \and
        V.~Dobrovolskas\inst{1}
        \and
        P.~Bonifacio\inst{2}
       }

\offprints{A.~Ku\v{c}inskas}

\institute{
        Institute of Theoretical Physics and Astronomy, Vilnius University, A. Go\v {s}tauto 12, Vilnius LT-01108, Lithuania \\
        \email{arunas.kucinskas@tfai.vu.lt}
        \and
        GEPI, Observatoire de Paris, CNRS, Universit\'{e} Paris Diderot, Place Jules Janssen, 92190 Meudon, France
}

\date{Received: date; accepted: date}

\abstract
{It is generally accepted today that Galactic globular clusters (GGCs) consist of at least two generations of stars that are different in their chemical composition and perhaps age. However, knowledge about the kinematical properties of these stellar generations, which may provide important information for constraining evolutionary scenarios of the GGCs, is still limited.}
{We study the connections between chemical and kinematical properties of different stellar generations in the Galactic globular cluster 47~Tuc.}
{To achieve this goal, we used abundances of Li, O, and Na determined in 101 main sequence turn-off (TO) stars with the aid of 3D hydrodynamical model atmospheres and NLTE abundance analysis methodology. We divided our sample TO stars into three groups according to their position in the \liona\ -- \naoo\ plane to study their spatial distribution and kinematical properties.}
{We find that there are statistically significant radial dependencies of lithium and oxygen abundances, \ALi\ and \AO, as well as that of \liona\ abundance ratio.  Our results show that first-generation stars are less centrally concentrated and dynamically hotter than stars belonging to subsequent generations. We also find a significant correlation between the velocity dispersion and O and Na abundance, and between the velocity dispersion and the \naoo\ abundance ratio.}
{}

\authorrunning{Ku\v{c}inskas et al.}
\titlerunning{47 Tuc: new ties between chemical and dynamical evolution of globular clusters}

\maketitle

\section{Introduction}

Results of numerous photometric and spectroscopic studies of Galactic globular clusters (GGCs) carried out during the past decade have provided strong observational evidence that most (if not all) GGCs consist of at least two stellar populations that are characterized by different chemical composition and, possibly, age \citep[see reviews by, e.g.,][and references therein]{M11,GCB12}. This may be a result of extended star formation that has taken place in the GGCs, which has led to the enrichment of the second-generation stars with light elements, such as Li, Na, Al, formed in the interiors of the first-generation stars. The common consensus today is that either AGB stars \citep[e.g.,][]{DDC12} or fast-rotating massive stars \citep[e.g.,][]{CCD13} may be held responsible for this enrichment, although alternative proposals exist as well (pollution by massive binaries, \citealt[][]{MPL09}; novae, \citealt[][]{MZ12}; or hypotheses that do not involve multiple populations, e.g., \citealt[][]{BLM13}). Nevertheless, current observational evidence does not yet allow us to clearly distinguish between the different proposed scenarios.

$N$-body simulations predict that stars belonging to different generations may have different spatial distributions and kinematical properties \citep[e.g.,][]{B10,B11,VMD13}. Recent photometric and spectroscopic studies of GGCs confirm that second-generation stars are more centrally concentrated, which seems to be a general property of the GGCs studied so far (e.g., \citealt[][]{GCB12}; for results on 47~Tuc see \citealt[][]{NGP11,MPB12,CPJ14}).

However, our knowledge about the kinematical properties of different stellar generations is still rather limited. A first systematic attempt to investigate connections between the chemical and kinematical properties of stars in 18 GGCs was carried out by \citet[][]{BBC12}. The authors concluded that there was no statistically significant relation between the sodium abundance and either velocity dispersion or systemic rotation of the cluster stars. The only exceptions were NGC~6388, 6441, and 2808, which showed hints of a possible dependence of the velocity dispersion on the Na abundance. In a more recent study of 47~Tuc, \citet[][]{RHA13} have found that MS stars with the bluest photometric colors (which have been identified by the authors as second-generation stars) showed a significant anisotropy in the proper motions along the radial direction. No such anisotropy was found for the first-generation stars. To our knowledge, these are the only observational facts that point to kinematical differences between the different stellar generations in GGCs. Clearly, more observational constraints are needed, thus we here focus on a more detailed investigation of the kinematical properties of chemically tagged MS stars in 47~Tuc.

\begin{figure}[tb]
\centering
\includegraphics[width=6cm]{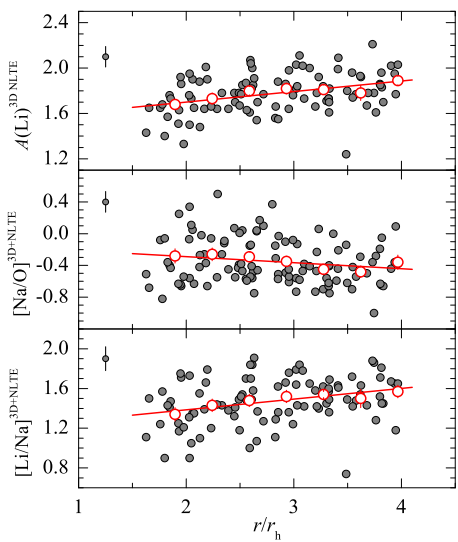}
\caption
{Lithium abundance, \ALi\ (top panel), \naoo, and \liona\ abundance ratios (middle and bottom panels) in our TO stars in 47~Tuc, plotted versus radial distance from the cluster center, $r/r_{\rm h}$ ($r_{\rm h}=174^{\prime\prime}$ is the half-mass radius of 47~Tuc, \citealt[][]{TDG93}). Solid lines are linear best fits to the data in the full sample of 101 stars. Large open circles are averages obtained in $1^{\prime}$ wide bins.
}
\label{fig:rad-distr}
\end{figure}

\section{Observational data\label{sect:observations}}

We used abundances of lithium, oxygen, and sodium that were determined by \citet[][]{DKB14} using the VLT GIRAFFE spectra of 101 main sequence turn-off (TO) stars in 47~Tuc. The investigated stars span a very narrow range in their atmospheric parameters (with peak-to-peak values of $\Delta \Teff \leq 310$~K and $\Delta \log g \leq 0.16$~dex), thus minimizing the possible influence of uncertainties in their effective temperatures and gravities on the abundance determination. Abundances of O and Na were determined using a 3D+NLTE methodology, where 1D~NLTE abundances were obtained using 1D hydrostatic \ATLAS\ model atmospheres, with the O and Na line profiles computed with the \MULTI\ code \citep{C86} in the implementation of \citet{KAL99}. 1D~NLTE abundances of O and Na were then corrected for the 3D hydrodynamical effects by adding 3D--1D~LTE abundance corrections computed using 3D hydrodynamical \COBOLD\ \citep[][]{FSL12} and 1D hydrostatic \LHD\ \citep[][]{CLS08} model atmospheres taken from the CIFIST 3D hydrodynamical model atmosphere grid \citep{LCS09}. To determine the abundance corrections, 3D/1D~LTE spectral line profiles of oxygen and sodium were computed with the spectral synthesis code \Linfor\footnote{http://www.aip.de/$\sim$mst/Linfor3D/linfor\_3D\_manual.pdf}.
The obtained 3D--1D~LTE abundance corrections were generally small, spanning the range of $0.02\dots0.09$~dex for O and $-0.04\dots0.03$~dex for Na, and very similar for all stars, thus their influence on the shape of possible abundance (anti-)correlations was negligible.

Abundances of lithium were determined using measured equivalent widths of the \ion{Li}{i} resonance line located at $670.8$~nm, which were then fed to the interpolation formula of \citet[][]{SBC10} to compute 3D~NLTE lithium abundances. Typical errors in the derived abundances of lithium, oxygen, and sodium were 0.09, 0.10, and 0.08~dex, respectively. Radial velocities of individual stars were determined using the IRAF task \textit{fxcorr}
\citep[see][for details]{DKB14}.

\begin{figure}[tb]
\centering
\includegraphics[width=7.5cm]{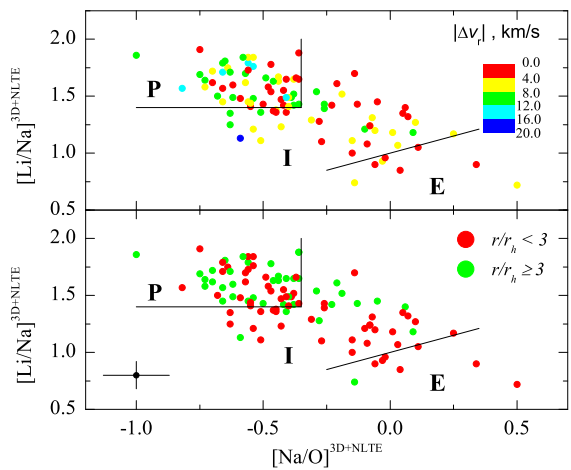}
\caption
{Distribution of TO stars in the $\liona-\naoo$ plane. Stars are color-coded according to (a) $|\Delta v_{\rm r}|$, which is the absolute radial
velocity of a given star in the reference system of the cluster (top panel); and (b) distance of each individual star from the cluster center, $r/r_{\rm h}$ (bottom panel). Solid lines delineate boundaries used to divide stars into groups P, I, and E (see Sect.~\ref{sect:discuss} for details).}
\label{fig:abund-ratio}
\end{figure}

\section{Results and discussion\label{sect:discuss}}

Radial distributions of lithium abundance, \ALi, as well as [Na/O] and [Li/Na] abundance ratios are plotted in Fig.~\ref{fig:rad-distr}. Linear regression analysis (performed taking errors on the $y$-axis into account) suggests the existence of weak but statistically significant radial dependencies of \ALi\ and \liona, with the probability that the two linear relations are real (computed using Student's $t$-test) equal to $p=0.9995$ and 0.996, respectively (similar results were obtained if errors on the $y$-axis were ignored). For \naoo, we obtained $p=0.92$. Although not shown in Fig.~\ref{fig:rad-distr}, a statistically significant relation was obtained also for \AO, with $p=0.97$. Similarly, \lioo\ and \ANa\ showed hints of the radial dependencies as well, although at a lower statistical significance ($p=0.85$ and 0.73).

For further analysis, we divided our TO star sample into three groups according to their position in the $\liona-\naoo$ plane (delineated by the solid lines in Fig.~\ref{fig:abund-ratio}): (a) primordial, P, $\liona > 1.4$ and $\naoo <-0.35$; (b) extreme, E, $\liona<0.6\times\naoo+1.0$; and (c) intermediate, I, which contains all other stars that did not fall into either group P or E. In making this selection, we loosely followed the criteria used by \citet[][]{CBG09a} to ensure that group P contains stars with \naoo\ ratios falling within an interval of 0.3~dex from the minimum \naoo\ value found in our TO sample, excluding outliers (in our case, this was furthermore complemented with a similar requirement for the \liona\ ratios). Group E was designed to contain $\sim10\%$ of the sample stars with the lowest \liona\ and highest \naoo\ ratios, to match the typical numbers of group E stars found in this cluster \citep[cf.][]{CPJ14,MPB12}. The only major difference between our selection criteria and those of \citet[][]{CBG09a} and \citet[][]{CPJ14} was that in our case we also tried to take into account the kinematical information seen in Fig.~\ref{fig:abund-ratio}: our criteria ensure that group P contains nearly all Li-rich and Na-poor (O-rich) stars with the highest absolute radial velocities, $|\Delta v_{\rm r}|>8$~km/s ($|\Delta v_{\rm r}|\equiv|v_{\rm rad}-\langle v_{\rm rad}\rangle^{\rm clust}|$, where $v_{\rm rad}$ is radial velocity of the individual star and $\langle v_{\rm rad}\rangle^{\rm clust}$ is the mean radial velocity of the cluster). Velocities this high are confined to stars with the lowest \naoo\ and highest \liona\ ratios (Fig.~\ref{fig:abund-ratio}, top panel), which makes it a natural choice to assign them to a single group.

\begin{figure}[tb]
\centering
\includegraphics[width=8cm]{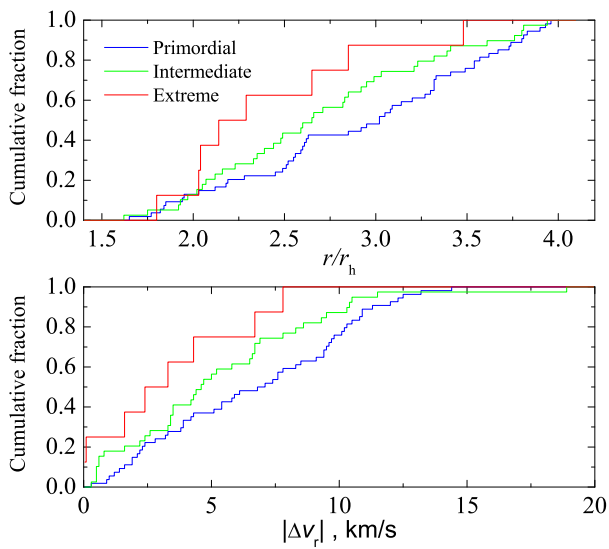}
\caption
{Cumulative number fraction of TO stars in the groups P, I, and E, plotted versus the distance from the cluster center (top panel) and absolute radial velocity, $|\Delta v_{\rm r}|$ (bottom panel).}
\label{fig:cumul}
\end{figure}

\begin{figure}[tb]
\centering
\includegraphics[width=8cm]{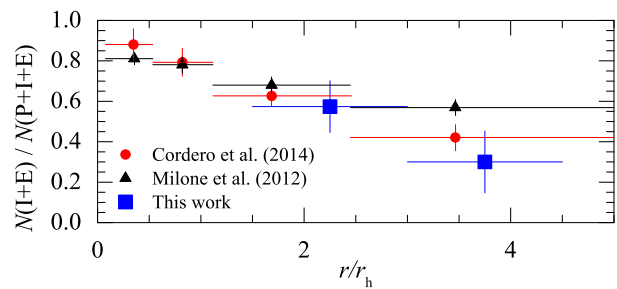}
\caption
{Ratios of stars in subsequent stellar populations in 47~Tuc, $N({\rm I+E})/N({\rm P+I+E})$, plotted versus distance from the cluster center. Data from the spectroscopic study of \citet[][]{CPJ14} are shown as red circles, photometric results of \citet[][]{MPB12} are plotted as black triangles, while our data are shown as large blue rectangles.}
\label{fig:nfrac}
\end{figure}

In fact, locations of the three groups in the $\naofe-\oofe$ plane (not shown here) are very similar to those of the corresponding P, I, and E groups of RGB stars identified in 47~Tuc according to their O and Na abundances by \citet[][]{CBG09a} and \citet[][]{CPJ14}. The main difference is in the resulting P:I ratio, which is significantly higher in our case (fractions of stars in group E are very similar), $\sim1.4$ versus $\sim0.4$ in \citet[][]{CBG09a} and $\sim0.7$ in \citet{CPJ14}. Despite this fact, the fraction of stars in the subsequent groups, $N({\rm I+E})/N({\rm P+I+E})$, obtained in our work agrees well with those determined by \citet[][see our Fig.~\ref{fig:nfrac} below]{CPJ14}. It may therefore be safe to assume that in all three studies groups P, I, and E trace the same stellar populations in 47~Tuc\footnote{Note that results obtained in our study remain unchanged if the selection criteria of \citet[][]{CBG09a} are used instead; the only difference in this case is that statistical significance of various relations between the chemical and kinematical properties of cluster stars becomes lower.}.

Visual inspection of Fig.~\ref{fig:abund-ratio} suggests that stars in groups I and E may be more concentrated toward the cluster center than stars in group P (Fig.~\ref{fig:abund-ratio}, bottom panel). Similar differences can be inferred if cumulative fractions of stars in groups P, I, and E are plotted versus distance from the cluster center (Fig.~\ref{fig:cumul}, top panel). Results of the Kolmogorov--Smirnov (K--S) test performed on the fractional distributions of stars plotted versus the radial distance from the cluster center (Fig.~\ref{fig:cumul}) show that the probability for the P and I groups to be drawn from the same population is low, $p=0.127$, with significantly lower probabilities for groups P--E and I--E ($p=6.0\times10^{-7}$ and $0.003$).

Information about the kinematical properties seen in Fig. \ref{fig:abund-ratio} (top panel) and \ref{fig:cumul} (bottom panel) also suggests that stars in group P have larger $|\Delta v_{\rm r}|$ compared to $|\Delta v_{\rm r}|$ in groups I and E. Indeed, a K--S test performed on the fractional distribution of stars with respect to their absolute radial velocities confirms that the probabilities for the P--I, P--E, and I--E groups to be drawn from the same populations are low and equal to $p=0.069$, $7\times10^{-7}$, and $1.7\times10^{-4}$, respectively.

We also find that the velocity dispersion in group E ($2.86\pm1.01$~km/s) is lower than that in group P ($3.84\pm0.52$~km/s): the probability of the two subsamples to be drawn from the same population is $p=0.055$ (obtained using the $F$-test); the probabilities for groups P--I and I--E are higher, $p=0.163$ and 0.125. Since our results also suggest that group E is more centrally concentrated, the two facts taken together would be difficult to explain using current theoretical models, which predict a higher velocity dispersion toward the cluster center. At the same time, we note that velocity dispersion calculated including stars in \textit{all} groups  indeed tends to increase toward the cluster center, in agreement with the earlier observational findings in 47~Tuc \citep[e.g.,][]{LBK10}.

\begin{figure}[tb]
\centering
\includegraphics[width=8cm]{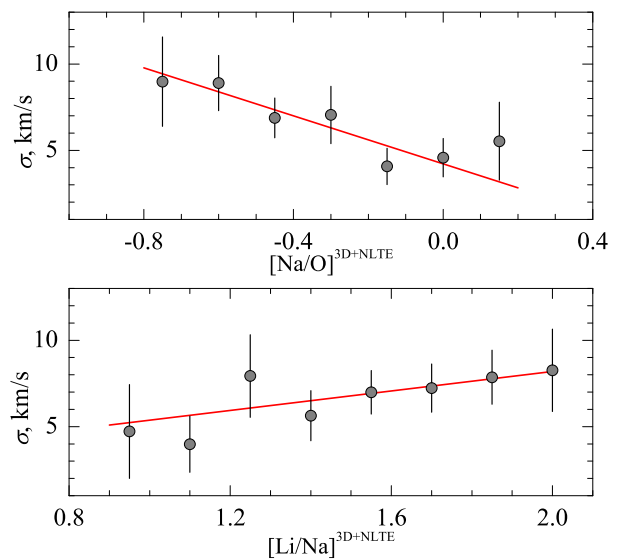}
\caption
{Velocity dispersion of TO stars plotted versus the \naoo\ (top panel) and \liona\ (bottom panel) abundance ratios. The red solid line is linear best fit to the data weighted according to the number of stars in each abundance bin.}
\label{fig:disper}
\end{figure}

\begin{figure}[tb]
\centering
\includegraphics[width=8cm]{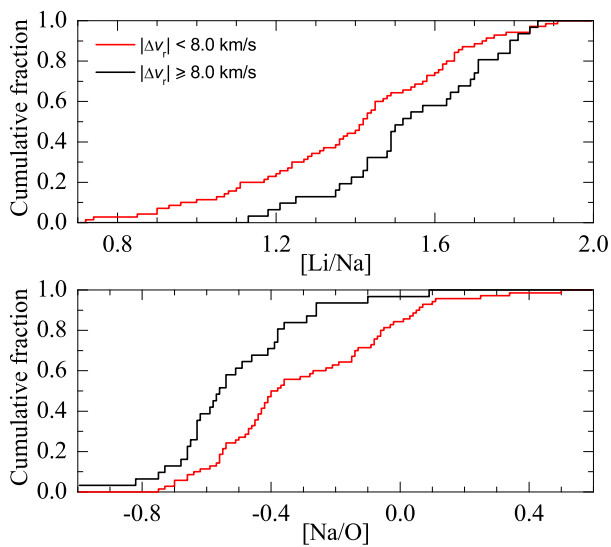}
\caption
{Cumulative number fraction of TO stars with $|\Delta v_{\rm r}|<8$~km/s and $|\Delta v_{\rm r}|\geq8$~km/s, plotted versus the \naoo\ (top panel) and \liona\ (bottom panel) abundance ratios.}
\label{fig:cumul-kin}
\end{figure}

It is important to stress that connections between the chemical and kinematical properties of TO stars are evident even without separating them into the three groups discussed above. This is seen in Fig.~\ref{fig:disper} where we plotted the velocity dispersion, $\sigma$, versus \naoo\ and \liona\ abundance ratios ($\sigma$ values were computed using all TO stars divided into 0.15~dex wide \naoo\ and \liona\ bins, allowing for 0.05~dex overlap between the bins). The probability that the linear relations (constructed using weights based on the number of stars in each abundance bin) are real is $p=0.996$ and 0.96 (Student's $t$-test) for the \naoo\ and \liona\ ratios (note that \citealt[][]{BBC12} also  detected hints of decreasing velocity dispersion with increasing sodium abundance in NGC~6388, 6441, and 2808). Significant dependencies are also seen when $\sigma$ is plotted versus the [O/Fe] and [Na/Fe] ratios (not shown here), with $p=0.991$ and 0.982. In addition, we divided all TO stars into two groups according to their velocities and assigned stars with $|\Delta v_{\rm r}|<8$~km/s to one group and those with $|\Delta v_{\rm r}|\geq8$~km/s to another (Fig.~\ref{fig:cumul-kin} shows the cumulative number fractions of stars in each group plotted versus the \naoo\ and \liona\ abundance ratios). The results of the K--S test show that the probability of the two kinematical groups to be drawn from the same population is low, with $p=6\times10^{-4}$ and 0.012 for the \naoo\ and \liona\ ratios.

Interestingly, \citet[][]{RHA13} have found in their study of proper motions of MS stars in 47~Tuc that more centrally concentrated MS stars have a clearly pronounced asymmetry in the proper motions along the radial direction. Unfortunately, we could not verify whether such signatures would also be detectable in radial velocities because of the limited sample size.

The ratios of subsequent stellar populations in a given cluster, $N({\rm I+E})/N({\rm P+I+E})$, may be useful in testing the existing theoretical evolutionary scenarios of the GGCs \citep[see, e.g.,][]{B10,B11,VMD13}. In this respect, the radial distribution of the number ratios $N({\rm I+E})/N({\rm P+I+E})$ determined in our study agrees very well with those of \citet[][]{CPJ14} and \citet[][]{MPB12}, which were obtained from the analysis of spectroscopic and photometric data, respectively (Fig.~\ref{fig:nfrac}).

\section{Summary and conclusions\label{sect:conclus}}

Our analysis of TO stars in 47~Tuc shows that there are statistically significant links between the radial distance of a given star from the cluster center and the abundance of lithium, oxygen, and sodium in its atmosphere. These radial relations are most clearly pronounced in the case of lithium and oxygen, with slightly lower statistical significance in the case of sodium. For the abundance ratios, a radial dependence is detected only for [Li/Na] with high statistical significance.

Our results also corroborate the existence of three stellar generations found earlier in this cluster by \citet[][]{CGB13} and \citet[][]{CPJ14}, although the fractions of stars in each generation derived in the latter two studies and our work are somewhat different, partly because of differences in the selection criteria and stellar samples used in the analysis. We find that the three generations may also differ in their spatial distributions and absolute radial velocities, $|\Delta v_{\rm r}|$, albeit at different levels of statistical significance. Our results suggest that stars belonging to the first generation have higher $|\Delta v_{\rm r}|$ values than those belonging to subsequent generations. At the same time, second- and third-generation stars are more centrally concentrated. We also find that radial velocity dispersion of stars in group E is lower than that in group P. Given that group E is more centrally concentrated than group P, the two facts may be difficult to explain simultaneously with the current evolutionary models, which predict that the velocity dispersion should increase toward the cluster center. We also detect a statistically significant dependence of velocity dispersion (calculated using all TO stars, i.e., without separating them into groups P, I, and E) on \naoo\ abundance ratio, with lower \naoo\ values associated with higher velocity dispersion. At the same time, there are significant differences in the distributions of the \naoo\ and \liona\ ratios in stars with $|\Delta v_{\rm r}|<8$~km/s and those with higher velocities.

Our results therefore suggest that 47~Tuc is not yet fully dynamically relaxed, which is in line with the predictions of $N$-body simulations \citep[e.g.,][]{VMD13}. On the other hand, 47~Tuc is the only GGC where direct connections between the chemical and kinematical properties of member stars could be detected so far. Obviously, it would very interesting to verify whether such connections are present in other GGCs as well, which would provide additional observational evidence for constraining possible evolutionary scenarios of the GGCs.

\begin{acknowledgements}

We thank the anonymous referee for useful comments that significantly helped to improve the paper. This work was supported by grants from the Research Council of Lithuania (MIP-065/2013) and the bilateral French-Lithuanian programme ``Gilibert'' (TAP~LZ~06/2013, Research Council of Lithuania).

\end{acknowledgements}

\bibliographystyle{aa}

\begin{thebibliography}{}

\bibitem[{Bastian} {et al.}(2013)]{BLM13}
Bastian,~N., Lamers,~H.J.G.L.M., de Mink,~S.E., et al.
2013, \mnras, 436, 2398

\bibitem[{Bekki} (2010)]{B10}
Bekki,~K.
2010, \apj, 724, L99

\bibitem[{Bekki} (2011)]{B11}
Bekki,~K.
2011, \mnras, 412, 2241

\bibitem[{Bellazzini} {et al.}(2012)]{BBC12}
Bellazzini,~M., Bragaglia,~A., Carretta,~E., et al.
2012, \aap, 538, A18

\bibitem[{Caffau} {et al.}(2008)]{CLS08}
Caffau,~E., Ludwig,~H.-G., Steffen,~M., et al.
2008, \aap, 488, 1031

\bibitem[{Carlsson} (1986)]{C86}
Carlsson,~M.
1986, UppOR, 33

\bibitem[{Carretta} {et al.}(2009)]{CBG09a}
Carretta,~E., Bragaglia,~A., Gratton,~R., et al.
2009, \aap, 505, 117

\bibitem[{Carretta} {et al.}(2013)]{CGB13}
Carretta,~E., Gratton,~R.G., Bragaglia,~A., D'Orazi,~V., \& Lucatello,~S.
2013, \aap, 550, A34

\bibitem[{Charbonnel} {et al.}(2013)]{CCD13}
Charbonnel,~C., Chantereau,~W., Decressin,~T., Meynet,~G., \& Schaerer,~D.
2013, \aap, 557, L17

\bibitem[{Cordero} {et al.}(2014)]{CPJ14}
Cordero,~M.J., Pilachowski,~C.A., Johnson,~C.I., et al.
2014, \apj, 780:94

\bibitem[{D'Antona} {et al.}(2012)]{DDC12}
D'Antona,~F., D'Ercole,~A., Carini,~R., Vesperini,~E., \& Ventura,~P.
2012, \mnras, 426, 1710

\bibitem[{de Mink} {et al.}(2009)]{MPL09}
de Mink,~S.E., Pols,~O.R., Langer,~N., \& Izzard,~R.G.
2009, \aap, 507, L1

\bibitem[{Dobrovolskas} {et al.}(2014)]{DKB14}
Dobrovolskas,~V., Ku\v{c}inskas,~A., Bonifacio,~P., et al.
2014, \aap, 565, A121

\bibitem[{Freytag} {et al.}(2012)]{FSL12}
Freytag,~B., Steffen,~M., Ludwig,~H.-G., et al.
2012, J. Comp. Phys., 231, 919

\bibitem[{Gratton} {et al.}(2012)]{GCB12}
Gratton,~R.G., Carretta,~E., \& Bragaglia,~A.
2012, \aapr, 20, 50

\bibitem[{Korotin} {et al.}(1999)]{KAL99}
Korotin, S. A., Andrievsky, S. M., \& Luck, R. E.
1999, \aap, 351, 168

\bibitem[{Lane} {et al.}(2010)]{LBK10}
Lane,~R.R., Brewer,~B.J., Kiss,~L.L., et al.
2010, \apj, 711, L122

\bibitem[{Ludwig} {et al.}(2009)]{LCS09}
Ludwig,~H.-G., Caffau,~E., Steffen,~M., et al.
2009, \memsai, 80, 711

\bibitem[Maccarone \& Zurek (2012)]{MZ12}
Maccarone,~T.J. \& Zurek,~D.R.
2012, \mnras, 423, 2

\bibitem[{Martell} (2011)]{M11}
Martell,~S.L.
2011, Astron. Nachr., 332, 467

\bibitem[{Milone} {et al.}(2012)]{MPB12}
Milone,~A.P., Piotto,~G., Bedin,~L.R., et al.
2012, \apj, 744, 58

\bibitem[{Nataf} {et al.}(2011)]{NGP11}
Nataf,~D.M., Gould,~A., Pinsonneault,~M.H., \& Stetsosn,~P.B.
2011, \apj, 736, 94

\bibitem[{Richer} {et al.}(2013)]{RHA13}
Richer,~H.B., Heyl,~J., Anderson,~J., et al.
2013, \apj, 771, L15

\bibitem[{Sbordone} {et al.}(2010)]{SBC10}
Sbordone,~L., Bonifacio,~P., Caffau,~E., et al.
2010, \aap, 522, 26

\bibitem[{Trager} {et al.}(1993)]{TDG93}
Trager,~S.C., Djorgovski,~S.G., \& King,~I. R.
1993, ASP Conf. Ser. 50, 347

\bibitem[{Vesperini} {et al.}(2013)]{VMD13}
Vesperini,~E., McMillan,~S.L.W., D'Antona,~F., \& D'Ercole,~A.
2013, \mnras, 429, 1913

\end{thebibliography}

\end{document}